\documentclass[
  twocolumn,
  prb,
  showpacs,
  amsmath,
  amssymb,
  superscriptaddress
]{revtex4}
\usepackage[colorlinks]{hyperref}
\usepackage{bm}
\usepackage{graphicx}

\newcommand{\pa}{\partial}

\newcommand{\be}{\begin{equation}}
\newcommand{\e}{\end{equation}}
\newcommand{\beml}{\begin{subequations}}
\newcommand{\eml}{\end{subequations}}
\newcommand{\beq}{\begin{eqnarray}}
\newcommand{\eq}{\end{eqnarray}}
\newcommand{\ba}{\begin{array}}
\newcommand{\ea}{\end{array}}
\newcommand{\bpm}{\begin{pmatrix}}
\newcommand{\epm}{\end{pmatrix}}
\newcommand{\bc}{\begin{cases}}
\newcommand{\ec}{\end{cases}}
\newcommand{\lt}{\left}
\newcommand{\rt}{\right}
\newcommand{\n}{\nonumber}
\newcommand{\la}{\langle}
\newcommand{\ra}{\rangle}

\newcommand{\h}{^\dagger}

\DeclareMathOperator{\tr}{Tr}

\DeclareMathOperator{\im}{Im}
\DeclareMathOperator{\re}{Re}

\begin{document}

\title{Korshunov instantons out of equilibrium}

\author{M.~Titov}
\affiliation{Radboud University, Institute for Molecules and Materials, NL-6525 AJ Nijmegen, The Netherlands}

\author{D.\,B.~Gutman}
\affiliation{Department of Physics, Bar Ilan University, Ramat Gan 52900, Israel}

\begin{abstract}
Zero-dimensional dissipative action possesses non-trivial minima known as Korshunov instantons. They have been known so far only for imaginary time representation that is limited to equilibrium systems. In this work we reconstruct and generalise Korshunov instantons using real-time Keldysh approach. This allows us to formulate the dissipative action theory for generic non-equilibrium conditions. Possible applications of the theory to transport in strongly biased quantum dots are discussed.
\end{abstract}
\date{\today}

\maketitle
\section{Introduction}
Quantum systems coupled to an environment have been extensively studied for many years following the seminal work by R.\,P.~Feynman and F.\,L.~Vernon.\cite{Feynman63} More recently a large variety of theoretical models that correspond to different physical realisations of dissipative systems have been proposed. These models include shunted Josephon junction,\cite{MSS}  tunnel junction,\cite{Zaikin_Schon} granular array,\cite{Altland2006} Luttinger liquid with a static impurity,\cite{Kane,Tsvelik} and open quantum dot.\cite{Matveev_Furusaki,Nazarov}  

Despite such a diversity the behaviour of these systems at low energies turned out to be similar. Indeed, at long times the systems become essentially zero dimensional and can be described using a collective degree of freedom that is often called the phase. In a quantum dot, for example, the phase can be viewed as a conjugate variable to the electron charge on the dot. The phase dynamics is non-local in time and is governed by Ambegaokar-Eckern-Sch\"on type of theory\cite{AES} (AES) that is sometimes referred to as dissipative action. 

Linearisation of the dissipative theory reproduces the so-called Caldeira-Leggett model\cite{Caldeira} that has a number of applications. In particular, the model is instrumental for analysing dephasing rate in q-bits\cite{Breuer} and dissipative decay rate of metastable states as well as for calculating other related quantities.\cite{Kamenev_book} It is worth mentioning that similar models were rigorously derived in the context of disordered low-dimensional systems as an effective theory for low-energy virtual fluctuations.\cite{GGM_virtual}

It has been realised already in Ref.~\onlinecite{Guinea86} that periodicity of the dissipative action with respect to the phase (which originates in the discrete nature of electron charge for the quantum dot model) may have some important consequences that are absent in the Caldeira-Leggett model. In field theory such non-perturbative effects are often accounted for by the so-called instanton solutions. For dissipative action those are Korshunov instantons\cite{Korshunov} that correspond to trajectories of the phase $\phi$ (a zero dimensional bosonic field) labeled by a winding number $W$. This number shows how many times the corresponding trajectory circles around the origin.

Instantons and their contribution to observable quantities have been thoroughly studied within Matsubara formalism that refers to quantum action which is taken on the imaginary time contour. In particular, the analysis of the action on the saddle point trajectory has been performed in Ref.~\onlinecite{Korshunov}, while the corresponding 't\,Hooft determinant\cite{tHooft} has been found in Refs.~\onlinecite{Zaikin_Panyukov,Grabert}. Later the two-loop renormalisation group approach to the problem has been developed\cite{Beloborodov} and the analogy with topological Pruisken instantons, which appear in topological sigma model with broken time reversal symmetry, has been established.\cite{Burmistrov}.  The equivalence of the AES model to the so-called ``paperclip'' model with $\theta$ topological term is demonstrated in Ref.~\onlinecite{Lukyanov}. Recent progress in the field is summarised in Ref.~\onlinecite{Shnirman15}.

Instantons distinctly manifest themselves in transport and thermodynamic measurements due to their topological nature. Instanton solutions in open quantum dots describe the emergence of Coulomb blockade\cite{Nazarov},   that manifest itself through weak modulation of the conductance with a gate voltage.  These are remains  of  Coulomb blockade  that dominates transport in  closed dots, where conductance  vanishes   except for  a set of narrow Coulomb peaks\cite{Glazman_review}.
In an open dot limit, the  Coulomb blockade effect is strongly suppressed,      
and results only  in  oscillations of the conductance with  a gate voltage. 
In this case the instantons with a certain winding number are responsible for the corresponding harmonics of conductance oscillation. This effect was  observed in recent experiments with controlled coupling between quantum dot and the leads.\cite{Liora2015}  
\begin{figure}
\includegraphics[width=\columnwidth]{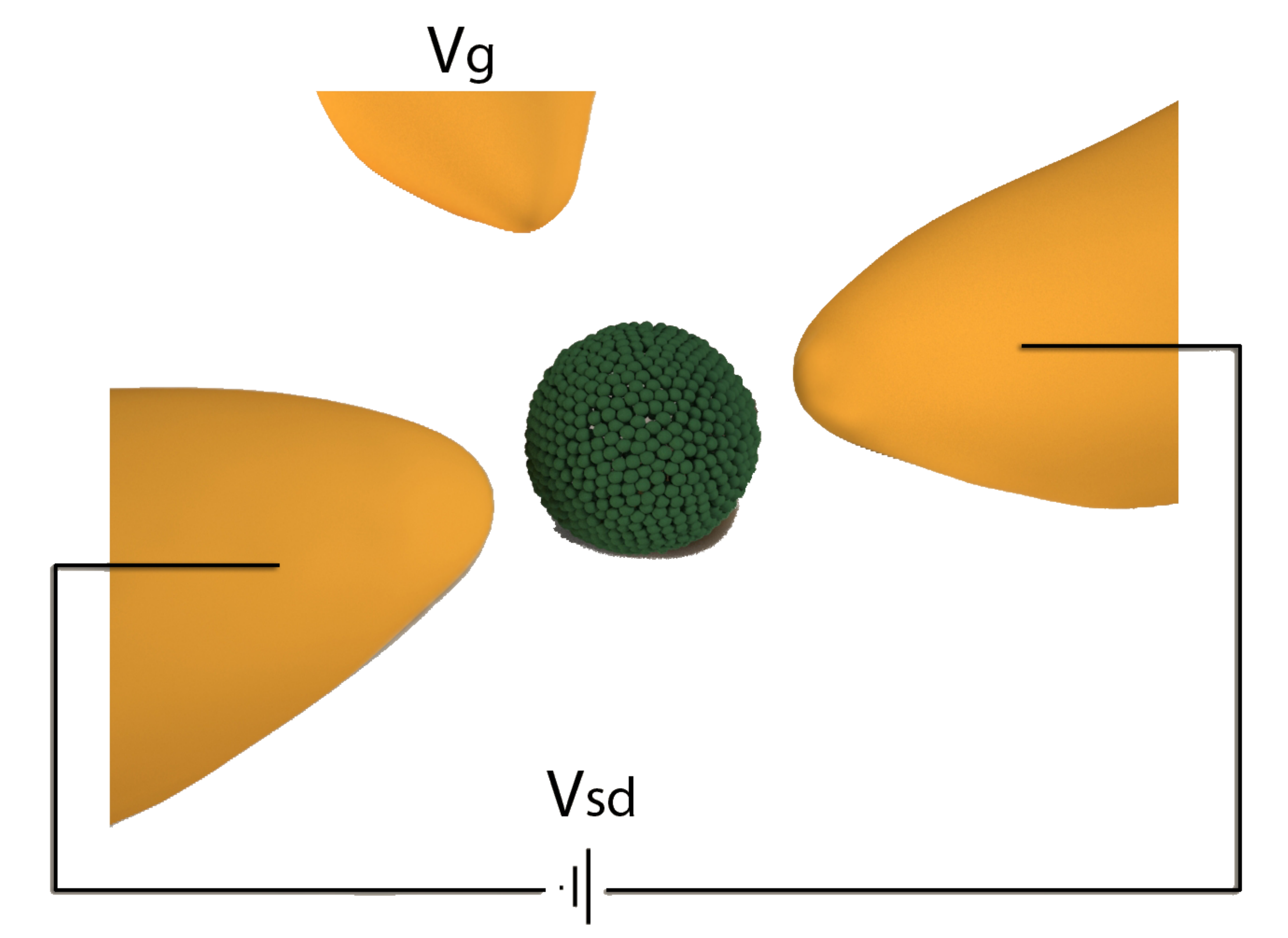}
\caption{Schematic  illustration of the setup: quantum dot, source,  drain and gate
electrodes.}
\label{fig1}
\end{figure}

Even though the physics of Coulomb blockade at equilibrium is fairly well understood its modification under generic non-equilibrium conditions remains  largely unexplored (see, however, Ref.~\onlinecite{Shnirman15}). Apart from clear experimental motivation there also exists a substantial theoretical interest to the problem. Non-perturbative analysis applied to systems which range from simple quantum mechanics to advanced field theory models is equivalent to finding ``classical trajectories'' that minimise corresponding imaginary time action.  In contrast, non-equilibrium conditions require the use of real time Keldysh formalism that may not reconcile with the non-perturbative equilibrium approach. 

Some progress in this direction has been achieved within a phenomenological framework utilising quantum Langevin equation.\cite{Golubev_Zaikin1997} While this approach captures many qualitative properties of the system it can be justified only in the topologically trivial sector while the consideration of instanton contributions requires more careful analysis. 
 
In this paper a non-perturbative approach based on Keldysh formalism is developed for a generic zero-dimensional non-linear theory.  A non-equilibrium extension of the dissipative action for a biased quantum dot is constructed and the corresponding integral saddle-point equations are derived and solved explicitly in terms of hypergeometric functions. For equilibrium conditions known results are reproduced. 

In the next Section we shell focus on the formulation of dissipative action in real time using an open quantum dot as a prime example.

\section{Dissipative action in real time}

Let us focus on a system consisting of a quantum dot that is schematically represented in Fig.~\ref{fig1}. The electron-electron interactions on the dot are described by the Hamiltonian $H_c=E_c (\hat{n}-n_g)^2$, where $\hat{n}$ is the electron density operator and $n_g$ is a background density set by the potential of the gate.  The characteristic charging energy $E_c=e^2/2C$ is determined by the capacitance $C$ of the dot and by the electron charge $e$.  We assume that the dot is metallic, with a single particle level spacing $\Delta$ that is smaller than all other energy scales in the problem. Each of the two leads coupled to the dot are assumed to be kept at thermodynamic equilibrium. The coupling is fully characterised by a unitary scattering matrix $S(E)$, which defines all transport characteristics of the dot at an energy $E$ in the absence of interactions.
 
Coulomb interaction on the dot is decoupled by Hubbard-Stratonovich  transformation.  As a result electrons on the dot are subject to the confining potentials, disorder and time dependent auxiliary bosonic filed. The problem at hand can be, therefore, viewed as the Ladauer-type transport problem in the presence of external quantum field. Alternatively, the interaction on the dot (that is taken in the form of the charging energy) can be placed on the leads. This transforms the problem to a network of interacting quantum wires (scattering channels)\cite{Bagrets} connected via a non-interacting quantum dot described by the scattering matrix $S$. 

Quite generally all physical observables in the system are given by the fermionic path integral 
\be
\langle \mathcal{O} \rangle=\int \mathcal{D}\Psi \mathcal{D}\bar{\Psi}\;\mathcal{O}[\Psi]\,e^{i\mathcal{A}[\Psi]}, 
\e
where the real-time fermonic action on the Keldysh contour $\mathcal{A}[\Psi] = \oint_K dt\,\lt(i\bar{\Psi}\pa_t \Psi-H\rt)$ can be decomposed to the free fermion and to the interaction parts
\be
\mathcal{A}[\Psi] =\mathcal{A}_0[\Psi]+\mathcal{A}_{\rm c}[\Psi].
\e
Following the standard route we shall perform the Hubbard-Stratonovich transformation to decouple interaction term by means of the bosonic field $\phi$ and integrate out the fermionic degrees of freedom. The resulting theory is given by the functional integral 
\be
\langle \mathcal{O} \rangle =\int \mathcal{D}\phi\; \mathcal{O}[\phi]\, e^{i A[\phi]}
\e
where $\dot{\phi}\equiv \pa_t\phi$ is a potential on the dot, which is uniform in space by construction. The resulting action is readily decomposed into a sum
\be
A[\phi]=A_{\rm c}[\phi]+A_{\rm dot}[\phi],
\e
that consists of the charging contribution
\be
\label{Ac}
A_{\rm c}[\phi]=-\frac{1}{2E_c}\oint_K\!\! dt\; \dot{\phi}^2(t) -n_g\oint_K\!\! dt\; \dot{\phi}(t),
\e
and the coupling term
\be
\label{Alog}
A_{\rm dot}=-i\ln Z[\phi].
\e
The latter is given by the sum of ``vacuum loops'' and can be generally written as
\be
\label{action_general}
Z[\phi]=e^{-i\dot{\phi}_c \Pi^a\dot{\phi}_q}\det\big[1-\hat{f}+\hat{R}[\phi]\hat{f}\big]\,
\e
where $\Pi^a$  is an advanced component of the fermionic polarisation operator, $\hat{f}$ is a single particle distribution function operator, which is diagonal in energy representation for stationary systems, and $\phi_c$($\phi_q$) is the classical (quantum) component of the bosonic field. Namely, $\phi_{c}=(\phi_++\phi_-)/2$ and $\phi_{q}=\phi_+-\phi_-$, where $\phi_{+(-)}$ stands for the bosonic field on the upper (lower) branch of the Keldysh contour. The operator $R$ is given by the potential dependent single-particle scattering matrix 
\be
R[\phi]=S^\dagger[\phi_+]S[\phi_-].
\e
In general the operator $R$ is non-local in time domain.

An essential simplification to the general theory of Eqs.~(\ref{Ac}-\ref{action_general}) is achieved in the so-called tunnelling limit, which corresponds to a weak coupling between the dot and the leads. The dependence of the scattering matrix on external potential $\dot{\phi}$ yields in the tunnelling limit
\be
S [\phi]=\bpm 
\hat{r} & \hat{t}'\, e^{i\phi}\\
\hat{t}\,e^{-i\phi} & \hat{r}'
\epm,
\e
where $\hat{r}(\hat{r}')$ and $\hat{t}(\hat{t}')$ are the matrices of reflection and transmission amplitudes in channel space for a non-interacting system. (Here we ignored the dependence of transmission and reflection probabilities on energy within the window given by $E_c$).

All transmission probabilities $T_n$, which are the eigenvalues of the matrix $\hat{t}\hat{t}\h$ (or $\hat{t}'(\hat{t}')\h$), are small in the tunnelling limit $T_n\ll 1$. Thus, it is legitimate to expand the logarithm in Eq.~\eqref{Alog} to the lowest order with respect to $\hat{t}$  and $\hat{t}'$. With the help of the identity $\ln\det \hat{L} = \tr\ln {\hat L}$ we obtain in the tunnelling limit 
\be
\label{atild}
A_{\rm dot}=\frac{g}{4}\int_{-\infty}^\infty \!\!\!
dt_1\,dt_2\, \Phi_L^T(t_1) \tilde{\alpha}(t_1,t_2) \Phi_R(t_2),
\e
where $g=g_L+g_R$, and  $g_{\eta}=\sum_n T_n^{\eta}$   is  the dimensionless  
coupling between the dot and the left ($\eta =L$) and the right ($\eta=R$)  leads,
also known as Caildera-Legget ``viscosity'' and  we  focus on the limit of a strongly coupled dot
 ($g \gg 1$);    $\Phi_{a}$ are the vectors in Keldysh space 
\be 
\Phi_L=\bpm z_+^{-1} \\ z_-^{-1} \epm, \quad \Phi_R=\bpm z_+\\  z_- \epm,\quad z=e^{-i \phi},
\e
where the index $\pm$ denotes the upper (lower) part of the Keldysh contour and the upper index $T$ stands for the vector transposition.
The current through   the quantum dot, can be written as
\begin{equation}
I=\frac{g_Lg_R}{g_L+g_R}\int d\epsilon[n(\epsilon-V/2)-n(\epsilon+V/2)]\nu(\epsilon)/\nu_0
\label{current}
\end{equation}
Here $n(\epsilon)$ is a Fermi-Dirac distribution function, $\nu(\epsilon)$ is a tunneling density of states, $g_{L/R}$ are couplings between the dot and the left and  the right leads. The tunneling density of states can be recast in terms of the bosonic fields as\cite{Altland2009}
\begin{align}
\nu(\epsilon)=&\nu_0\,\re\int d\tau e^{i \epsilon\tau}\bigg[
(1-n(\tau))\lt\la e^{i\phi_-(\tau)-i\phi_+(0)}\rt\ra\n \\
&+n(\tau)\lt\la e^{i\phi_+(\tau)-i\phi_-(0)}\rt\ra\bigg]\,,
\end{align}
where $\nu_0$ is the density of states for non-interacting electrons.
The problem is thus reduced to the calculation of the tunneling density of states
for this system, both in and out of equilibrium. 
\subsection{thermal equilibrium}
In thermal equilibrium the matrix integration kernel in Eq.~\eqref{atild} becomes the function of a time difference only, $\tilde{\alpha}=\tilde{\alpha}(t_1-t_2)$, where
\be
\tilde{\alpha}(t)=\bpm \tilde{\alpha}_{++} & \tilde{\alpha}_{+-} \\ \tilde{\alpha}_{-+} & \tilde{\alpha}_{--} \epm
\e
is parameterised by the following functions of time 
\beml
\label{kernel}
\beq
&&\tilde{\alpha}_{++}(t)=\tilde{\alpha}_{--}(t)=-i \re \lt[\frac{T^2}{\sinh^2(\pi Tt+i0)}\rt],\qquad\\
&&\tilde{\alpha}_{+-}(t)=-\tilde{\alpha}_{-+}^*(t)= \frac{iT^2}{\sinh^2(\pi T t-i0)},
\eq
\eml
where $T$ is the temperature.

\begin{figure}
\includegraphics[width=\columnwidth]{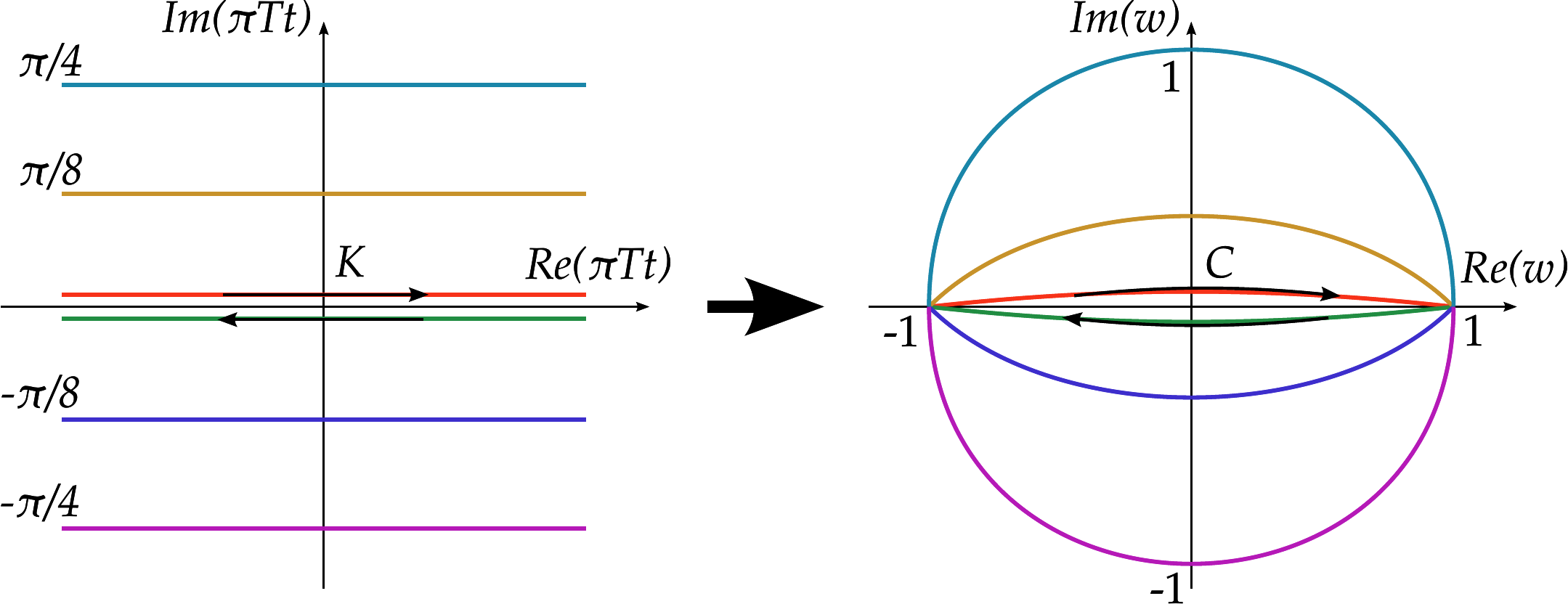}
\caption{Schematic illustration of the conformal map $w$ in the time complex plane, which maps the Keldysh contour K onto the contour C.}
\label{fig2}
\end{figure}

Analytic properties of the kernel $\tilde{\alpha}(t)$ in Eq.~(\ref{kernel}) are dictated by the shape of the Keldysh contour $K$, which goes around a cut directed along the real time axis. One may reformulate the problem by taking advantage of a conformal map of the time complex plane onto the complex plane of the parameter $w$\cite{Braunecker,Muzykanskii} 
\be
\label{w}
w=\tanh(\pi T t).
\e
Under such transformation the strip in the complex plane $-i/4T<\im t < i /4T$ is mapped onto the interior of the unit circle $|w|<1$ as shown in Fig.\,\ref{fig2}. Consequently, the Keldysh contour $K$ is transformed into the contour $C$ that surrounds a cut $w\in (-1,1)$ in the complex $w$ plane.

In terms of the variable $w$ the theory is described by the action 
\be
\label{action_w}
A_{\rm dot}=g\int_{-1}^{1}\!\frac{dw}{2\pi} \int_{-1}^{1}\!\frac{dw'}{2\pi}\,\Phi_L^T(w)\alpha(w-w')\Phi_R(w').
\e
The kernel of the integral transform $\alpha(w)$ has the following components   
\beml
\label{kern2}
\beq
&&\alpha_{++}=\alpha_{--}=-\frac{i}{2}
\lt[\frac{1}{(w+i\delta)^2}+\frac{1}{(w-i\delta)^2}\rt],\qquad\\
&&\alpha_{+-}(w)=\alpha_{-+}(-w)=\frac{i}{(w+i\delta)^2}.
\eq
\eml
where we introduce a finite shift $\delta$ which implies that the corresponding pole is situated outside the countur $C$. The functions $z_{\pm}(w)=z(w \pm i 0)$ are related to a complex function $z(w)$ that is analytic everywere in the complex $w$-plane with the exception of the cut $w\in (-1,1)$.

One can note that taking the limit of zero temperature in the kernel $\tilde{\alpha}$ given by Eq.~\eqref{kernel} would lead to the same expressions as in Eq.~\eqref{kern2} but with the cut extending over the entire real axis. Thus the only difference between zero and finite temperature analysis is encoded in terms of the variable $w$ in analytical properties of the function $z(w)$. 

The action (\ref{action_w}) gives rise to the following saddle-point equations on the analytic function $z(w)$,
\beq
&&\oint_Cdw' \lt[\frac{z_\pm(w)}{z(w')}-\frac{z(w')}{z_\pm(w)}\rt]\n\\ 
&&\times \lt(\frac{1}{(w-w')^2} +\frac{1}{(w-w'\pm i\delta)^2}\rt)=0\,.
\label{saddle_w}
\eq
Here  the integral over  the running variable was closed into a countur $C$,  such that the pole at $w'=w$ lays inside the contour $C$, while the pole at $w'=w\pm i\delta$ is outside the integration contour.

Equations \eqref{saddle_w} were derived by minimizing the action $A_\textrm{dot}$ only. Taking into account the charging part of the action $A_c$ gives rise to the terms which are subleading in the parameter $T/gE_c$ as shown in the Appendix \ref{Matsubara}). 
At this level of accuracy, i.e.\,for $T/gE_c \ll 1$, the subleading terms have to be disregarded.

Despite the approximations made the saddle-point equations \eqref{saddle_w} are the non-linear integral equations. The existing mathematical methods are, however, limited to the linear equations only.\cite{Muskelishvili,Stone} Fortunately, the restrictive analytic structure of Eqs.~(\ref{saddle_w}) makes further analysis possible. 

It is instructive to take advantage of the analogy with the Matsubara representation, which corresponds to the contour $C$ taken as the unit circle. Using analytic properties of the kernel $\alpha$ as described in Appendix \ref{transform} one can demonstrate that 
\be 
\label{instanton}
z_1(w)=\frac{w-\xi}{1-\bar{\xi}w}
\e 
is an instanton trajectory with a winding number equal to one, where the parameters $\xi$ and 
$\bar{\xi}$ are independent real numbers on the cut $\xi,\bar{\xi} \in (-1,1)$. The solution of Eq.~\eqref{instanton} minimizes the action \eqref{action_w}. 

The value of the action (\ref{action_w}) on the intanton trajectory is readily evaluated as 
\be
\label{Adot1}
A_{\rm dot}[z_1]=i g/2,
\e
while the charging part of the action evaluated on the same trajectory gives
\be 
\label{Ac1}
A_{\rm c}[z_1]=2i\pi^2\frac{T}{E_c}\frac{\bar{\xi}-\xi}{1-\bar{\xi}\xi}+n_g.
\e 
The results of Eqs.~(\ref{Adot1},\ref{Ac1}) are fully equivalent to those obtained in Matsubara representation in the Appendix~\ref{Matsubara}. 

Similarly the instanton solution, which corresponds to a winding number $W$, is given by 
\be 
\label{zW}
z_{W}(w)=\prod_{i=1}^W\frac{w-\xi_i}{1-\bar{\xi_i}w},
\e 
with the real parameters $\xi_i,\bar{\xi}_i \in (-1,1)$. Substituting Eq.~\eqref{zW} in to the action (\ref{action_w}) one finds
\be 
\label{result_W}
A_{\rm dot}[z_W]=i g W/2,
\e 
while the charging part of the action is still of the order of $T/E_c$ corresponding to a generalization of Eq.~\eqref{Ac1}. The charging contribution is irrelevant as far as $T/gE_c \ll 1$. 
By expanding Eq.(\ref{current})  over the instantons in the low voltage limit, one finds  a d.c. 
conductance
\be
\label{conductance_instanton}
G \propto G_{\rm Drude}\bigg[1+\sum_W e^{-g|W|/2}\cos(2\pi Wn_g)\bigg]\,,
\e
where
\be
G_{\rm Drude}=\frac{e^2}{2\pi}\frac{g_Lg_R}{g_L+g_R}\,.
\e
Note, that the contribution of the instantons with a winding number $W$ lead to the 
appearance of $W$ of the d.c. conductance with the gate voltage.
Such harmonics were recently observed experimentally\cite{Liora2015}.  
The fluctuations  around the saddle point, account for the zero bias anomaly, leading to 
the renormalziation of the coupling strength $g \rightarrow g+\ln T/E_c$, \cite{Glazman2005,Burmistrov,Beloborodov}, and  consequently  to
a temperature dependent prefactors in Eq.(\ref{conductance_instanton}).

Thus, the construction of real-time Korshunov instanons using Keldysh formalism at equilibrium is analogous to that in Matsubara representation once the conformal map (\ref{w}) is employed (see Appendix~\ref{Matsubara} for technical details). The apparent advantage of the Keldysh approach is due to the possibility to generalize it for a generic non-equilibrium situation.

\subsection{beyond equilibrium}

To be more specific we shall focus below on the case of a quantum dot, which is strongly biased by a source-drain voltage $V$. In this case, the single-particle distribution function inside the dot can be parameterized as
\be
f(\epsilon)=\frac{g_L}{g}f_0(\epsilon_-)+\frac{g_R}{g}f(\epsilon_+),
\e
where $\epsilon_-=\epsilon-eVg_R/g$ and $\epsilon_+=\epsilon+eVg_L/g$.

In terms of the variable $w$ the integral kernel in Eq.~(\ref{action_w}), which depends now separately on two arguments, is given by
\begin{align}
\label{kernel_neq}
&\alpha_{ij}(w,w')=\alpha^{(0)}_{ij}(w-w')\;L\!\lt(\frac{1-w}{1+w}\frac{1+w'}{1-w'}\rt),\\
&L(x)= \frac{g_L^2+g_R^2}{g^2}+\frac{g_Lg_R}{g^2}\lt[x^{i\nu}+x^{-i\nu}\rt],\quad
\label{Lfunction}
\end{align}
where $\nu=eV/2\pi T$ and $\alpha^{(0)}_{ij}(w)$ is the equilibrium kernel set by Eq.~\eqref{kern2}. The corresponding saddle-point equations on the analytic function $z(w)$ reads 
\be
\oint_C\! dw' \lt[\alpha(w_\pm,w')\frac{z(w')}{z_\pm(w)}- \alpha(w',w_\pm)\frac{z_\pm(w)}{z(w')}\rt]=0,\quad
\label{saddle_point_general}
\e
where $z_\pm(w)=z(w\pm i0)$, $w_\pm=w\pm i\delta$, and $w\in (-1,1)$.  

We shell look for the solution of Eq.(\ref{saddle_point_general}) using the following Ansatz
\be
\label{ansatz}
z(w)=\frac{w-\xi}{1-\bar{\xi}w}F(w)\,,
\e
which is an obvious extension of the equilibrium instanton of Eq.(\ref{instanton}). Motivated by the analytic properties of the solution at equilibrium and kernel of integration (\ref{kernel}), we shell require that the functions $F(w)$ and $1/F(w)$ have no singularities outside the contour $C$. That is equivalent to the requirement that all zeroes and singularities of $F(w)$ are confined to the interval $w\in (-1,1)$. Under such an assumption the saddle point equations can be easily solved (see Appendix~\ref{non_eq} for the details). 

A general instanton solution with a winding number one is given by an indefinite integral
\be
\label{non_equilibrium_instanton}
z(w)=(1-\xi\bar{\xi})\int\frac{dw}{(1-\bar{\xi}w)^2}\,L\!\lt(\frac{1-w}{1+w}\frac{\bar{\xi}+1}{\bar{\xi}-1}\rt),
\e
that can be explicitly taken as
\begin{align}
z(w)& =K\lt[ \frac{w-\xi}{1-w\bar{\xi}} + R + 2\frac{g_Lg_R}{g^2}\frac{1-\xi\bar{\xi}}{1-\bar{\xi}}\rt.\n\\
 &\lt.\times\lt(\mathcal{B}_\nu(x)+\mathcal{B}_{-\nu}(x)+\frac{(1-w)(1+\bar{\xi})}{1-w\bar{\xi}}\rt)\rt],\quad
\label{final0}
\end{align}
where $K=F(1/\bar{\xi})$, $R$ is an integration constant and the following definitions have been used
\begin{align}
&x=\frac{1-w}{1+w}\frac{\bar{\xi}+1}{\bar{\xi}-1},\\
&\mathcal{B}_\nu(x)=x^{i\nu}\lt(_1F_2(1,i\nu;1+i\nu;x)-\frac{1}{1-x}\rt).
\end{align}
Here, $_1F_2(a,b;c;x)$ is the hypergeometric function.

It is worth noticing that the integration constant $R$ can be absorbed in the redefinition of the parameters $\xi$ and $K$ by means of the transformation $\xi'=(\xi-R)/(1-R\bar{\xi})$ and $K'=K(1-R\bar{\xi})$. By performing this transformation we effectively set $R=0$ in Eq.~(\ref{final0}). The remaining parameter $K$ reflects the overall scaling invariance of the saddle-point equation (\ref{saddle_point_general}). Its choice does affect neither physical observables nor required analytic properties of the function $F(w)$. Hence we can set $K=1$ without loss of generality. 

Thus, the resulting instanton solution is given by Eq.~(\ref{final0}) with $K=1$ and $R=0$. Using known properties of the hypergeometric function we can indeed check that the solution obtained corresponds to $F(w)$ having analytic properties required by the construction. For a finite value of voltage the instanton solution acquires a branch cut in the interval $(-1,1)$. In the limit of equilibrium (for $g_L=0$, or $\nu=0$) the solution reduces to a familiar form of Eq.~(\ref{instanton}).

Since this interval $w\in(-1,1)$ precisely coincides with an image of the Keldysh contour on $w$ plane one may calculate the value of the action on the saddle point solution (\ref{ansatz}), by closing the integrals in the exterior of this interval in the complex plane. As a result one gets  
\begin{equation}
A_{\rm dot}[z_1]=\frac{g}{4\pi} \oint_C d \ln z(w_1)=\frac{ig}{2}\,.
\end{equation}
Thus, though both, the action and the saddle point solution, are affected by the external bias, the value of the action on the saddle point trajectory out of equilibrium remains the same. The instanton solutions with a higher winding numbers may be constructed in a similar fashion.
 
\section{conclusions and outlook}

In this work we have studied the real-time dissipative action. We established that the real-time Keldysh and imaginary-time Matsubara framework are almost identical in  equilibrium once a conformal map of the time plane into $w$-plane is employed. Topological instantons that solve integral saddle-point equations for the Keldysh action are constructed.  In equilibrium the solutions are represented by slightly distorted M\"obius maps of $w$ Riemann sphere onto itself. The index of the map is equal to the winding number of instanton. 

The real-time theory is, however, further extended to the case of generic non-equilibrium situation. The saddle-point equations in non-equilibrium are derived and solved analytically to construct the non-equilibrium extension of Korshunov instantons. In this paper we have focused in particular on the interacting quantum dot that  is pushed out of equilibrium by applying a large {\em dc} bias voltage. We demonstrated that, in this case, the saddle point solution acquires the branch cut along the entire Keldysh contour. Remarkably, the modified instanton solution of a dissipative action survives such a generic non-equilibrium condition and  the value of the action on the instanton trajectory remains unchanged. It implies, for instance, that the conductance oscillations with multiple periods (\ref{conductance_instanton}) emerge when a finite voltage bias is applied.

Our results may be relevant for a large number of physical situation that are modelled by the dissipative action. To make a more detailed prediction for observable quantities 
such as temperature and voltage dependence of electric current via non-equilibrium interacting quantum dot,  would require the computation of fluctuation-determinant in the vicinity of the instanton solution obtained. Such an analysis is beyond the scope of the present paper and is relegated to a separate publication.

This work has been supported by ISF (grant 584/14), GIF (grant 1167-165.14/2011), the Dutch Science Foundation NWO/FOM 13PR3118, and by the EU Network FP7-PEOPLE-2013-IRSES Grant No 612624 ``InterNoM''. We acknowledge useful discussions with R.~Berkovits, L.~Bitton, I.~Burmistrov, A.~Frydman, I.~Gornyi, A.~Kamenev, A.~Mirlin, R.~Santos, and E.~Dalla Torre.

\appendix

\section{Korshunov Instantons in Matsubara representation}
\label{Matsubara}

The dissipative action in the Matsubara representation can be written as
\beq
\mathcal{S} &=&\frac{g}{4}\int_0^\beta\!\!\! d\tau_1d\tau_2\,\tilde{\alpha}(\tau_{12})\,e^{i\phi(\tau_1)-i\phi(\tau_2)}\n\\
&&+\frac{1}{4E_c}\int_0^\beta\!\! d\tau\,\dot{\phi}^2-in_g\int_0^\beta\!\! d\tau\,\dot{\phi},
\label{Saction}
\eq
where $\beta=1/T$ and $\tau$ stands for imaginary time. The integration kernel here reads
\be
\tilde{\alpha}(\tau)=\frac{T^2}{2}\lt[\frac{1}{\sin^2 (\pi T\tau+i0)}+\frac{1}{\sin^2 (\pi T \tau-i0)}\rt].
\e
In order to simplify the comparison with the results obtained in the main text we define the variable
\be
u=e^{2\pi i T\tau},
\e
which is used to rewrite the action (\ref{Saction}) in the form
\beq
\mathcal{S}&=&\frac{g T^2}{2}\oint_{O} \frac{du_1}{2\pi i u_1}\,\frac{du_2}{2\pi i u_2}\, e^{i\phi(u_1)-i\phi(u_2)}\alpha(u_1/u_2)\n\\
&&+\frac{i\pi T}{2E_c}\oint_O du u \left(\frac{\partial \phi}{\partial u}\right)^2- i n_g \oint_O du \frac{\partial \phi}{\partial u},
\label{Su}
\eq
where the integration is taken over the unit circle $O$ that corresponds to $|u|=1$ and the transformed kernel is defined as
\be
\alpha(u)=\frac{u}{(u_+-1)^2}+\frac{u}{(u_--1)^2},
\e
with $u_\pm=u (1\pm \delta)$, $\delta <1$.

Minimising the action \eqref{Su}  one derives the saddle point equation
\beq
&&\frac{8\pi T}{gE_c}\frac{\partial}{\partial u} u\frac{\partial \phi}{\partial u}+ \oint_{O} du' \lt(e^{i\phi(u')-i\phi(u)}-e^{i\phi(u)-i\phi(u')}\rt)\n\quad\\
&&\times\lt(\frac{1}{(u'-u_-)^2}+\frac{1}{(u'-u_+)^2} \rt)=0,
\eq
which can be solved following the original work by Korshunov.\cite{Korshunov}
The solution $\phi=\phi_1(u)$, which corresponds to an instanton with the winding number one, reads 
\be
\label{instanton_Matsubaro}
e^{i\phi_1}\simeq c_1\frac{u-z_1}{u-\zeta_2}\lt(1-\frac{2T}{gE_c}\lt[\frac{\zeta_2}{(u-\zeta_2)^2}-
\frac{\zeta_1}{(u-\zeta_1)^2}\rt]\rt),
\e
where we neglect the terms of the order of $(T/gE_c)^2$. The result of Eq.~(\ref{instanton_Matsubaro}) is parameterised by a constant $c_1$ and 
complex numbers $\zeta_{1,2}$ such that $|\zeta_1| >1$ and $|\zeta_2|<1$. For $T/gE_c\ll 1$ the solution represents a slightly distorted M\"obius transform.

One may further require that the absolute value of the bosonic field is fixed hence the phase $\phi$ is real. This requirement corresponds to further constraint
$\zeta_2=1/\zeta_1^*$ and $c_1=\zeta_2e^{i\lambda}$ with a real $\lambda$. Under such a constraint the result of Eq.~(\ref{instanton_Matsubaro}) represents a map of the interior of the unit circle to itself. Even though the additional constraint on the field seems reasonable we have not been able to justify it with a formal argument. 

The leading part of Eq.~(\ref{instanton_Matsubaro}) describes a circle in the complex plane that encompasses the origin just once. The sub-leading terms deform the circle without affecting its topology. In the other words, the function still corresponds to a closed contour that encompasses the origin once. The terms of the order $T/gE_c$ are usually neglected in literature since they do not affect the value of the action at the saddle point solution. 

The action evaluated on the instanton trajectory of Eq.~\eqref{instanton_Matsubaro} is given by
\begin{equation}
\mathcal{S}[e^{i\phi_1}]=\frac{g}{2}+\frac{\pi^2 T}{4E_c}\frac{\zeta_1+\zeta_2}{\zeta_1-\zeta_2},
\end{equation}
thus reproducing the results of Eqs.~(\ref{Adot1},\ref{Ac1}) of the main text for the real-time action $A=i\mathcal{S}$.

Similarly one can construct the instanton trajectory $\phi=\phi_W(u)$ with the winding number $W$ larger than one,  
\be
\label{Winst}
e^{i\phi_W(u)}=c_W\prod_{i=1}^W\frac{u-\zeta_1^i}{u-\zeta_2^i},
\e
which is parameterised by an arbitrary number $c_W$ and $2W$ complex coordinates such that $|\zeta_1^i|>1$, $|\zeta_2^i|<1$.
The result of Eq.~\eqref{Winst} corresponds to the action
\be
\label{Swres}
\mathcal{S}[e^{i\phi_W}]=\frac{gW}{2}+\frac{\pi^2 T}{4E_c}\sum_{i,j=1}^W\frac{\zeta_1^i+\zeta_2^j}{\zeta_1^i-\zeta_2^j}.
\e
The leading term in Eq.~(\ref{Swres}) is again manifestly equivalent to the result of Eq.~\eqref{result_W} of the main text. 

\section{integral transform}
\label{transform}

In this appendix we shall summarise some of the integrals which were used in the main text. An integral transform with Cauchy type kernel defined on the contour $C$ leads to the following results
\beml
\beq
\oint_C \frac{dw_1}{2\pi i}\, \frac{w_1^n}{w_1-w-i\delta} &=&
\bc
0,& n > 0 \\
-w^n, &  n<0
\ec \\
\oint_C \frac{dw_1}{2\pi i}\,\frac{w_1^n}{w_1-w} &=&
\bc
w^n, &  n > 0 \\
0, & n<0
\ec
\eq
\eml
For $z=z_1(w)$ given by the instanton trajectory of Eq.~\eqref{instanton} one further obtains
\beml
\beq
&&\oint_C\frac{dw'}{2\pi i}\, \frac{z^{-1}(w')}{(w'-w)^2}=0,\\
&&\oint_C\frac{dw'}{2\pi i}\, \frac{z(w')}{(w'-w)^2} =\frac{1-\xi\bar{\xi}}{(1-\bar{\xi}w)^2},\\
&&\oint_C\frac{dw'}{2\pi i}\, \frac{z(w')}{(w'-w+i\delta)^2}=0,\\
&&\oint_C\frac{dw'}{2\pi i}\, \frac{z^{-1}(w')}{(w'-w+i\delta)^2}=\frac{1-\xi\bar{\xi}}{(1-\bar{\xi}w)^2},
\eq
\eml
hence one can easily check that the solution (\ref{instanton}) satisfies the saddle point equation of Eq.~(\ref{saddle_w}).

\section{saddle-point equation out of equilibrium}
\label{non_eq}
The saddle-point equations (\ref{saddle_w}) for the analytic function $z(w)$, which reduces to $z_\pm(w)$ on the upper (lower) side of the brunch cut $w\in(-1,1)$, can be written as
\be
\label{basic}
z^2(w)I_1[z](w)=I_2[z](w),
\e
where we defined the integrals
\beml
\begin{align}
I_1[z](w)=&\oint_C dw'\,\lt(\frac{1}{(w-w')^2}+\frac{1}{(w-w'+i\delta)^2} \rt)\n\\
&\times L\!\lt(\frac{1-w}{1+w}\frac{1+w'}{1-w'}\rt)\,\frac{1}{z(w')},\\
I_2[z](w)=&\oint_C dw'\,\lt(\frac{1}{(w-w')^2}+\frac{1}{(w-w'+i\delta)^2} \rt)\n\\
&\times L\!\lt(\frac{1-w}{1+w}\frac{1+w'}{1-w'}\rt)\,z(w'),
\end{align}
\eml
where the pole at $w'=w$ lays inside the contour $C$ while the pole at $w'=w-i\delta$ is outside $C$ and the limit $\delta\to 0$ is assumed. The function $L(x)$ is defined in Eq.~(\ref{Lfunction}).
With the help of the Ansatz of Eq.~(\ref{ansatz}) supplemented with the assumption on analytic properties of the function $F(w)$ (such that all zeroes and singularities of $F(w)$ are confined to the interior of the contour $C$) one finds 
\beml
\label{eq_i}
\begin{align}
I_1=&-2\pi i \frac{\partial}{\partial w'}\lt[ L\!\lt(\frac{1-w}{1+w}\frac{1+w'}{1-w'}\rt) \frac{1}{z(w')}\rt]_{w'\to w},\\
I_2=&-2\pi i\lt( \frac{\partial}{\partial w'}\lt[ L\!\lt(\frac{1-w}{1+w}\frac{1+w'}{1-w'}\rt) z(w')\rt]_{w'\to w}\rt.\n\\
&+\lt.2\frac{\eta(\xi-\eta) F(\eta)}{(w-\eta)^2} L\!\lt(\frac{1-w}{1+w}\frac{1+\eta}{1-\eta}\rt) \rt),
\end{align}
\eml
where we introduced the parameter $\eta=1/\bar{\xi}$ for convenience. With the help Eqs.~(\ref{eq_i}) we readily rewrite Eq.~(\ref{basic}) in the trivial form
\be
\label{nice}
\frac{\pa z(w)}{\pa w}=\eta(\eta-\xi)F(\eta)  L\!\lt(\frac{1-w}{1+w}\frac{1+\eta}{1-\eta}\rt),
\e
that is solved by Eq.~(\ref{final0}). The following indefinite table integral has been used for the integration of Eq.~(\ref{nice})
\begin{align}
&\int \frac{dw}{(w-\eta)^2} \lt(\frac{1-w}{1+w}\frac{1+\eta}{1-\eta}\rt)^{i\nu}\n\\
&\qquad\qquad = \frac{2}{\eta^2-1} \mathcal{B}_\nu\!\lt(\frac{1-w}{1+w}\frac{1+\eta}{1-\eta}\rt)+\textrm{const},
\end{align}
where we defined 
\be
\mathcal{B}_\nu(x)= x^{i\nu}\lt(\frac{1}{1-x}-  \,_1F_2(1,i\nu;1+i\nu;x)\rt).
\e
The result of integration in Eq.~(\ref{nice}) is, then, written as
\begin{align}
&z(w)=C+\frac{\eta(\xi-\eta)F(\eta)}{w-\eta}\lt(\frac{g_L^2+g_R^2}{g^2}+2\frac{g_L g_R}{g^2}\frac{w-\eta}{\eta^2-1}\rt.\n\\
&\times \lt.\lt[\mathcal{B}_\nu\!\lt(\frac{1-w}{1+w}\frac{1+\eta}{1-\eta}\rt)+\mathcal{B}_{-\nu}\!\lt(\frac{1-w}{1+w}\frac{1+\eta}{1-\eta}\rt)\rt]\rt).
\end{align}
To obtain the result in the form of Eq.~(\ref{final0}) we have redefined the integration constant $C$ as
\be
C=K\lt(-\eta+2 \frac{g_Lg_R}{g^2} \frac{\eta(\eta-\xi)}{\eta-1}\rt)+KR,
\e
where $K=F(\eta)$ and $R$ is a new integration constant.

\end{document}